\documentclass[aps,prd,notitlepage,twocolumn,nofootinbib,amsmath,amssymb,eqsecnum]{revtex4-2}
\usepackage[utf8]{inputenc}
\usepackage[T1]{fontenc}
\usepackage{mathtools}
\usepackage{upgreek}
\usepackage{graphicx}
\usepackage{dcolumn}
\usepackage{bm}
\usepackage{dsfont} 
\usepackage[colorlinks=true,allcolors=teal]{hyperref}
\usepackage{tabularx}
\usepackage{epstopdf}
\usepackage{tensor}
\usepackage{mathrsfs}
\usepackage[normalem]{ulem}
\usepackage[usenames]{color}
\usepackage[dvipsnames]{xcolor}
\usepackage{soul}
\usepackage{cleveref}

\newcommand\be{\begin{equation}}
\newcommand\ba{\begin{eqnarray}}
\newcommand\ee{\end{equation}}
\newcommand\ea{\end{eqnarray}}
\newcommand\bw{\begin{widetext}}
\newcommand\ew{\end{widetext}}

\makeatletter
\newcommand{\labeltext}[3][]{%
    \@bsphack%
    \csname phantomsection\endcsname% in case hyperref is used
    \def\tst{#1}%
    \def\refmarkup{}%
    \ifx\tst\empty\def\@currentlabel{\refmarkup{#2}}{\label{#3}}%
    \else\def\@currentlabel{\refmarkup{#1}}{\label{#3}}\fi%
    \@esphack%
    \labelmarkup{#2}% visible printed text.
}
\makeatother

\newcommand{\pde}[2]{\frac{\partial{#1}}{\partial{#2}}}

\newcommand{\UVA}{Department of Physics, University of Virginia, P.O.~Box 400714, Charlottesville, Virginia 22904-4714, USA}

\begin{document}

\title{Probing dark-matter effects with gravitational waves using the \\ parameterized post-Einsteinian framework}

 \author{Eileen Wilcox}
 \affiliation{\UVA}

 \author{David A.~Nichols}
 \email{david.nichols@virginia.edu}
 \affiliation{\UVA}
 
 \author{Kent Yagi}
\email{kyagi@virginia.edu}
 \affiliation{\UVA}

\date{\today}

\begin{abstract} 
A massive black hole can develop a dark-matter overdensity, and the dark matter changes the evolution of a stellar-mass compact object inspiraling around the massive black hole through the dense dark-matter environment.
Specifically, dynamical friction speeds up the inspiral of the compact object and causes feedback on the dark-matter distribution. 
These intermediate mass-ratio inspirals with dark matter are a source of gravitational waves (GWs), and the waves can dephase significantly from an equivalent system in vacuum.
Prior work has shown that this dephasing needs to be modeled to detect the GWs from these systems with LISA (the Laser Interferometer Space Antenna); it also showed that the density and distribution of dark matter can be inferred from a GW measurement. 
In this paper, we study whether the parametrized post-Einsteinian (ppE) framework can be used to infer the presence of dark matter in these systems.
We confirm that if vacuum waveform templates are used to model the GWs from an inspiral in a dark-matter halo, then the resulting parameter estimation is biased.
We then apply the ppE framework to determine whether it can reduce the parameter-estimation biases, and we find that adding one ppE phase term to a waveform template eliminates the parameter-estimation biases (statistical errors become larger than the systematic ones), but the effective post-Newtonian order in the ppE framework must be specified without uncertainties.
When the post-Newtonian order has uncertainty, we find that the systematic errors on the ppE and the binary's parameters exceed the statistical errors. 
Thus, the simplest ppE framework would not give unbiased results for these systems, and a further extension of it, or dedicated parameter estimation with gravitational waveforms that include dark-matter effects would be needed. 
\end{abstract}

\maketitle

\tableofcontents

\section{Introduction} \label{sec:intro}

There is compelling evidence for the presence of dark matter (DM) over a wide range of length scales, and observations indicate that it is abundant, cold and weakly interacting~\cite{Bertone:2016nfn}.
The Planck satellite, for example, determined that dark matter accounts for 26.8\% of the mass-energy content in the universe~\cite{Planck:2018vyg}.
There have been extensive experimental and observational programs to detect the particles that compose dark matter, but given the feeble interactions between it and baryonic matter or its small self-interactions, these efforts have not yet determined the fundamental constituents of dark matter (see, e.g., the reviews~\cite{Bertone:2004pz,Feng:2010gw,Gaskins:2016cha,Schumann:2019eaa}).
This has motivated not only improving the existing searches for dark matter, but also exploring new methods to determine its nature~\cite{Bertone:2018krk}.
One recent effort is using gravitational waves (GWs) to search for the presence of high densities of dark matter that can form around black holes~\cite{Bertone:2019irm}.

LIGO's successful detection of GWs in 2015~\cite{LIGOScientific:2016aoc} not only confirmed a long-standing prediction of Einstein's theory of relativity, but also created a new opportunity to study fundamental gravitational physics~\cite{LIGOScientific:2016lio,Yunes:2016jcc}.
Nearly one-hundred confident detections of mergers of compact  objects have now been made~\cite{LIGOScientific:2018mvr,LIGOScientific:2020ibl,KAGRA:2021vkt}, which have tested the predictions of general relativity precisely using an large battery of tests~\cite{LIGOScientific:2017adf,LIGOScientific:2021aug,LIGOScientific:2018dkp,LIGOScientific:2020tif,LIGOScientific:2021sio}.
The spectrum of GW observations has opened further with the likely detection of a stochastic GW background by pulsar timing arrays~\cite{NANOGrav:2023gor,Reardon:2023gzh,EPTA:2023fyk,Xu:2023wog}.
It will continue to expand in the next decade with next-generation GW detectors, such as the planned LISA~\cite{LISA:2017pwj} (in space) and Einstein Telescope~\cite{Sathyaprakash:2012jk,Maggiore:2019uih} or Cosmic Explorer~\cite{Evans:2021gyd}.
These new detectors will not only expand the GW frequency spectrum, but also provide new opportunities to study fundamental physics.

Our focus in this paper will be on the LISA detector and the possibility of using the GWs from the inspiral of a neutron star (NS) into an intermediate-mass black hole (IMBH) with a surrounding dense distribution of dark matter, as a means to learn about the surrounding dark-matter environment.
The inspiral of a compact object into an IMBH---known as an intermediate mass-ratio inspiral (IMRI)---or a similar inspiral with a supermassive BH primary (an extreme mass-ratio inspiral) are a key GW source class for the LISA detector.
These systems have been well studied in the absence of a surrounding DM distribution, and have been established to provide precise information about the spacetime geometry of the primary massive BH (see, e.g.,~\cite{Amaro-Seoane:2012lgq}).
A non-vacuum environment could alter the IMRI's rate of inspiral and affect the GWs emitted from these systems.
In fact, prior work has confirmed this: if dark matter is present around an IMBH in sufficiently large amounts, the density of a spherically symmetric DM overdensity (referred to as a DM ``spike''~\cite{Gondolo:1999ef}) could be inferred from a GW measurement with LISA~\cite{Eda:2013gg,Eda:2014kra}. 

Dark-matter particles scatter gravitationally with the NS as it inspirals, and there is a net transfer of energy from the NS to the DM spike; this effect, known as ``dynamical friction''~\cite{Chandrasekhar:1943ys,Chandrasekhar:1943ii,Chandrasekhar:1943iii}, causes the NS to inspiral into the IMBH more rapidly than an equivalent IMRI in vacuum would. 
Previous work by Eda et al.~\cite{Eda:2014kra} showed that dynamical friction led to a large change in the rate of inspiral that allowed the distribution of dark matter to be inferred precisely. 
For much of the parameter space of IMRIs with dark matter in~\cite{Eda:2014kra}, it was subsequently shown in~\cite{Kavanagh:2020cfn} that neglecting feedback from dynamical friction onto the DM distribution significantly overestimated the increase in the rate of inspiral. 
Reference~\cite{Kavanagh:2020cfn} introduced a formalism to evolve the DM spike that takes into account the energy imparted to the DM distribution through dynamical friction.
The IMRI's orbit and the dark matter distribution were evolved simultaneously as a coupled set of ordinary and partial differential equations over the longer timescales associated with gravitational radiation reaction and dynamical friction.
The DM density in the vicinity of the NS secondary in this dynamic case was transiently depleted during the inspiral, which decreased the effect of dynamical friction and the amount of dephasing from a vacuum IMRI with the same masses.
The GW dephasing was also a more nontrivial function of frequency in the dynamic case than in the static (non-evolving) case considered in~\cite{Eda:2014kra}.

To determine how well the IMRIs with a dynamic DM distribution could be studied by LISA, Ref.~\cite{Coogan:2021uqv} performed a simulation study of the prospects for measurement, detection, and distinction from vacuum IMRIs.
It was shown in~\cite{Coogan:2021uqv} that the detection horizon for IMRIs with DM had very little difference from vacuum binaries; however, vacuum IMRI waveforms were also not capable of fitting an IMRI with a dynamic DM distribution when optimized over the vacuum IMRI parameters.
Thus, Ref.~\cite{Coogan:2021uqv} created a phenomenological analytical model of the GW phase in the frequency domain to perform Bayesian parameter-estimation studies. 
In the two test cases considered in~\cite{Coogan:2021uqv}, the power-law that determines the radial fall-off of the initial DM density could be inferred to a few and a few tens of a percent, respectively.
This established that properties of the DM could still be inferred in the dynamic case via a GW measurement with LISA, and that the previous studies that assumed the DM was static during the inspiral significantly overestimated the precision to which the DM distribution could be measured.

Since presence of dark matter in an IMRI can produce a measurable GW dephasing from analogous vacuum systems, it is useful to know how precisely this dephasing needs to be modeled to be detectable and have minimal bias in the inferred GW parameters. 
While using the true signal is the optimal choice for matched-filtering-based data analysis, there are more model-agnostic frameworks, such as the parametrized post-Einsteinian (ppE) approach~\cite{Yunes:2009ke,Chatziioannou:2012rf,Tahura:2018zuq}, which aim to capture deviations from vacuum waveforms in general relativity (GR).
The ppE framework has been shown to detect deviations from GR in simulation studies~\cite{Cornish:2011ys,Sampson:2013lpa,Sampson:2013jpa} and also has been applied to the existing GW events to constrain such deviations~\cite{LIGOScientific:2016lio,Yunes:2016jcc,LIGOScientific:2018dkp,LIGOScientific:2020tif,LIGOScientific:2021sio}.
Parameterized waveforms similar to those used in the ppE framework have also been applied to study effects of the astrophysical environment in IMRIs, in particular those induced by dark matter (see in~\cite{Cardoso:2019rou,CanevaSantoro:2023aol,Rivera:2024xuv}).
However, neither of the signal models used in~\cite{Cardoso:2019rou,Rivera:2024xuv} were as complex as the phase in the dynamic case in~\cite{Coogan:2021uqv}, so it has not been established how well the ppE framework can capture effects induced by frequency-domain signals with a more complicated morphology. 

In this paper, we address this question by investigating how well the ppE framework can account for these dynamic DM-induced GW dephasing from vacuum binaries. 
As a first calculation, we study the amount of systematic bias in parameter estimation if the GW signal is an IMRI with a surrounding distribution of dark matter, but a GR template waveform is used to analyze the GW signals. 
Next, we use a simple ppE waveform to analyze the GWs from an IMRI with dark matter.
Our ppE model modifies the vacuum gravitational waveform phase in the frequency domain by the addition of a single term involving a power of GW frequency.
This term has a constant overall scaling $\beta$ and the exponent is denoted by $b$; the $b$ coefficient determines the post-Newtonian (PN) order of the ppE term.
We use a Fisher-matrix formalism to compute the statistical errors on waveform parameters and address whether the ppE template is able to capture the DM effects on the gravitational waveform. 

Throughout this paper we focus on IMRIs in which the secondary is a neutron star, and the dark matter has negligible interactions with the nuclear matter in the neutron star.
This assumption is necessary, because recent work~\cite{Nichols:2023ufs} showed that a black-hole secondary can accrete enough mass during the inspiral to induce a dephasing of hundreds of GW cycles with respect to systems in which the effects of only dynamical friction were modeled. 
It also leads to nontrivial changes in the distribution of dark matter after the merger.
Other work~\cite{Karydas:2024fcn} also generalized the prescription to IMRIs on eccentric orbits. 

We now summarize our main findings. 
We first determined that if we use the vacuum GR template waveform to analyze the GW signal from an IMRI in a DM spike, then vacuum parameters, such as the chirp mass, are significantly biased. 
This is consistent with the results in~\cite{Coogan:2021uqv}. 
We next studied these IMRIs using ppE templates. 
In this context, we first found the value of ppE exponent $b$ by minimizing the systematic error on the IMRI's ``vacuum'' (those unrelated to the DM) parameters.
We assumed perfect knowledge of $b$ (i.e., $b$ was fixed) and computed the uncertainties on the ppE parameter $\beta$. 
This procedure decreased the systematic bias on vacuum parameters below their systematic errors, and the systematic error on $\beta$ was much smaller than the statistical error, too. 
The systematic error on the vacuum parameters, however, is very sensitive to the choice of $b$, which is not expected to be known \emph{a priori} and requires fine tuning to minimize the systematic errors. 
Consequently, we also performed another analysis in which we compute uncertainties on both $\beta$ and $b$; in this context, we found that systematic errors are now larger than statistical errors. 
This suggests that the ppE template has the potential to remove the bias and measure the GW effects induced by the DM spike, but future work is necessary to determine if more general ppE waveforms with the larger parameter sets could improve the level of systematic and statistical errors.

The rest of the paper is organized as follows. 
Section~\ref{Section:GW_SetUp} explains the waveform models that we use for IMRIs in a dark-matter environment and for the ppE framework. 
In Sec.~\ref{Section:Fisher}, we review the Fisher framework that we use to compute statistical and systematic errors. 
Section~\ref{Section:Motivation_Tests} presents the main questions that we address in this paper and the methods we use to answer them.
Our main results are given in Sec.~\ref{Section:Results}, and they are followed by our conclusions in Sec.~\ref{Section:Conclusions}. 
We give a few supplemental results in two Appendixes.
We use the geometric units $c=G_N=1$ throughout this paper.

\section{Gravitational waves from a compact-binary inspiral in a dark-matter halo} \label{Section:GW_SetUp}

\begin{figure}[t]
    \centering
    \includegraphics[width=\columnwidth]{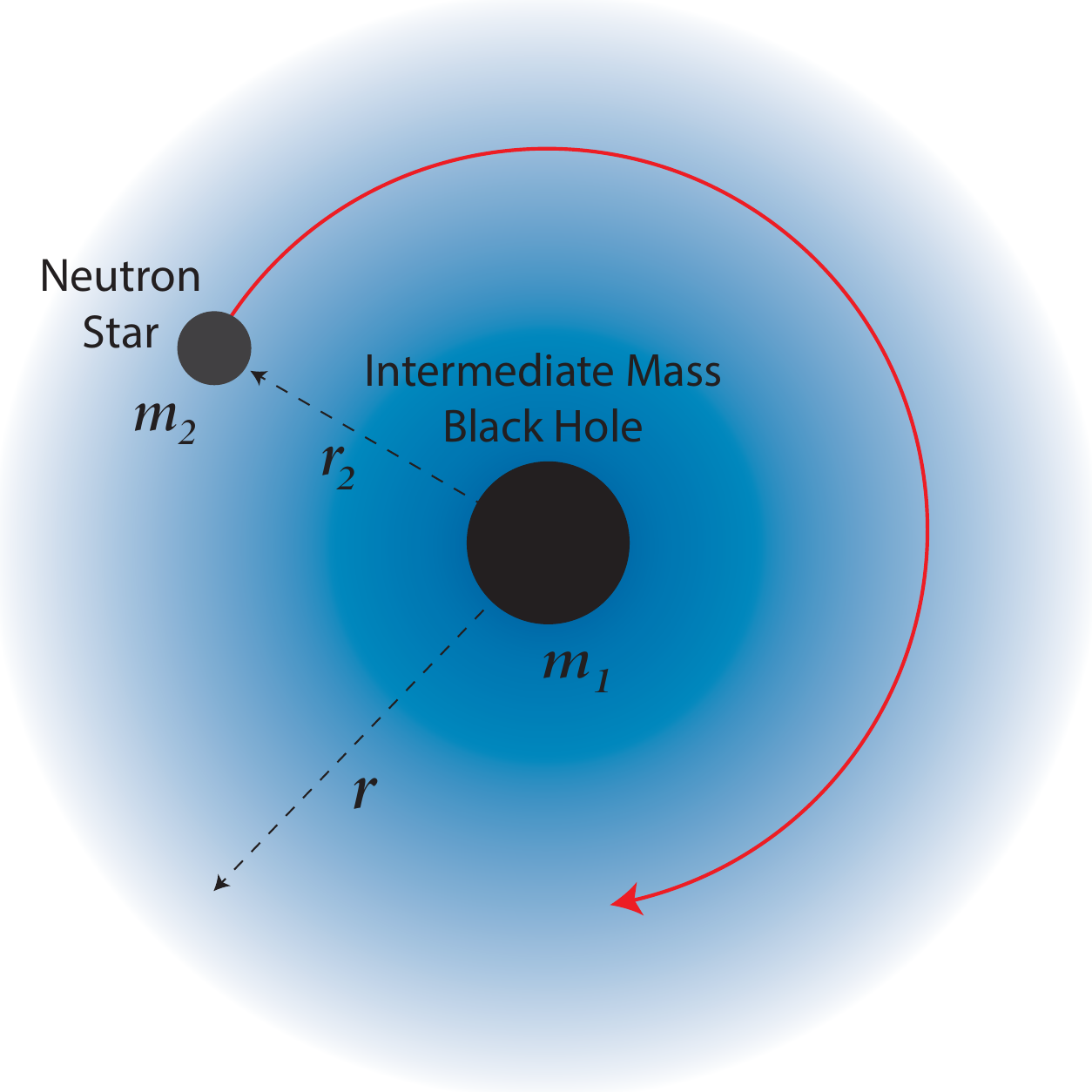}
    \caption{ 
    \textbf{Schematic of an IMRI with a DM spike}. 
    The IMBH has mass $m_1$ between $10^3$--$10^5$\,M$_\odot$ and is orbited by a neutron star of mass $m_2 = 1.4$\,M$_\odot$.
    The orbital radius is denoted by $r_2$, and $r$ is the distance from the center of the IMBH.}
    \label{fig:schematic}
\end{figure}

A schematic of the IMRIs that we will study in this paper is given in Fig.~\ref{fig:schematic}.
We will assume that the IMBH has a mass $m_1$ between $10^3$--$10^5 \, \text{M}_{\odot}$, and will be the primary (more massive object) in the binary.
The secondary (less massive object) will be a neutron star with mass $m_2 = 1.4 \, \mathrm{M}_\odot$.
Following Refs.~\cite{Kavanagh:2020cfn,Coogan:2021uqv}, we assume that the initial DM density around the IMBH is a spherically symmetric and isotropic spike given by 
\begin{equation} \label{eq:rho-init}
\rho_{\text{DM}}(r)= \begin{cases}
    \rho_{\text{sp}}(r_{\text{sp}}/r)^{\gamma_{\text{sp}}} & r_{\text{in}}\leq r\leq r_{\text{sp}} \\
    0 & r< r_{\text{in}}
\end{cases} .
\end{equation}
In Eq.~\eqref{eq:rho-init}, $r$ denotes the distance from the IMBH, $\gamma_\mathrm{sp}$ is the radial power law, $\rho_\mathrm{sp}$ is the normalization of the density, $r_\mathrm{in} = 4 m_1$ is the inner radius of the spike, and $r_{\text{sp}}$ is the outer radius of the spike.\footnote{At radii $r > r_\mathrm{sp}$, the spike will match onto the larger-scale DM halo, such as a Navarro-Frenk-White~\cite{Navarro:1995iw} profile; however, we will not treat the DM distribution at these larger radii in this paper.} 
Following~\cite{Eda:2014kra,Kavanagh:2020cfn}, we do not treat $r_\mathrm{sp}$ as an independent parameter, but as being determined by $r_\mathrm{sp} \approx 0.2 r_h$, where $r_h$ is defined by 
\begin{equation}
    \int_{r_\mathrm{in}}^{r_h} \rho_\mathrm{DM} (r) d^3r = 2m_1 .
\end{equation}
Throughout this paper, we will restrict to $\gamma_{\text{sp}} = 7/3$ and $\rho_{\text{sp}} = 226$ M$_\odot$/pc$^3$, which were the values used in the model of the spike in~\cite{Eda:2013gg,Eda:2014kra}.

As discussed in Sec.~\ref{sec:intro}, the evolution of this IMRI must be performed simultaneously with the evolution of the dark matter.
The \textsc{HaloFeedback} code~\cite{HaloFeedback} produces the orbital or GW phase from such systems, but a single numerical evolution takes sufficiently long that it is inefficient to use it in parameter-estimation studies.
Thus, Ref.~\cite{Coogan:2021uqv} constructed a phenomenological waveform model for the time-domain phase (as a function of GW frequency), which was calibrated over different mass-ratio IMRIs for a range of different $\gamma_\mathrm{sp}$ and $\rho_\mathrm{sp}$ values for the initial density in Eq.~\eqref{eq:rho-init}.
The phase is valid for IMRIs on circular orbits at Newtonian order; thus, it was used for just the dominant $l=2$, $m=2$ spherical-harmonic mode of the gravitational waveform.

We next review this phenomenological waveform, which we will use throughout the paper as the ``true'' GW signal.

\subsection{Gravitational-wave phase in the time domain}\label{Section:GW_T_Domain}

The ansatz for the time-domain phenomenological phase in~\cite{Coogan:2021uqv}, when written as a function of the GW frequency, $f$, is given by
\begin{align} \label{GeneralPhaseEq}
\Phi(f)  \equiv {} &  \Phi^V(f)\Big{\{}1-\eta y^{-\lambda}\nonumber\\
&\times\Big{[}1- \prescript{}{2}{\text{F}}_1\left(1,\vartheta, 1+\vartheta, -y^{-5/(3\vartheta)}\right)\Big{]}\Big{\}} . 
\end{align}
Here, $\prescript{}{2}{\text{F}}_1(a,b,c,z)$ is our notation for a Gaussian hypergeometric function, and the multiplicative factor $\Phi^\mathrm{V}(f)$ is the vacuum phase: 
\be \label{Vacuum Phase}
\Phi^V(f) = \frac{1}{16}(\pi \mathcal M f)^{-5/3} .
\ee
We used $\mathcal M=(m_1+m_2)\nu^{3/5}$ to denote the chirp mass, and $\nu =m_1m_2/(m_1+m_2)^2$ for the symmetric mass ratio. 
The variable $y$ is a dimensionless frequency, $y = f/f_t$, where $f_t$ is a constant frequency scale.

To describe the parameter choices that reproduce the GW phase from IMRIs in static and dynamic DM distributions, it is useful to define the following quantities. 
We first introduce a frequency scale $f_\mathrm{eq}$ defined by 
\begin{equation} \label{eq:feq}
    f_\mathrm{eq} = c_{f}^{3/(11-2\gamma_\text{sp})} .
\end{equation}
The coefficient $c_f$ is given by
\be \label{Eq:cf}
c_f = \frac{5}{8m_1^2}\pi^{2(\gamma_{\text{sp}}-4)/3}(m_1+m_2)^{(1-\gamma_{\text{sp}})/3}r_{\text{sp}}^{\gamma_{\text{sp}}}\xi \rho_{\text{sp}} \log \Lambda .
\ee
The new parameters introduced in this expression are $\log \Lambda$, the Coulomb logarithm, where the value of $\Lambda$ is given by $\Lambda = \sqrt{m_1/m_2}$ as in~\cite{Kavanagh:2020cfn}. 
The factor $\xi$ is the fraction of DM particles moving slower than the orbital speed of the secondary.
It was shown in~\cite{Nichols:2023ufs} to be given by
\be
\xi = 1-I_{1/2}\left(\gamma_{\text{sp}}-\frac{1}{2}, \frac{3}{2} \right) ,
\ee
where $I_{\frac{1}{2}}$ is the regularized incomplete beta function.
For $\gamma_\mathrm{sp} = 7/3$, this has the numerical value $\xi \approx 0.58$ used in~\cite{Kavanagh:2020cfn}.

The ansatz for the phase in Eq.~\eqref{GeneralPhaseEq} is sufficiently general that it can be used to describe the GW phase for both static DM distributions in~\cite{Eda:2014kra} and the dynamic ones in~\cite{Kavanagh:2020cfn,Coogan:2021uqv}, for different choices of the parameters $\eta$, $\lambda$, $\vartheta$ and $f_t$.
By choosing 
\be \label{eq:static_params}
\eta = 1 , \quad 
\lambda = 0 , \quad 
\vartheta = \frac{5}{11-2\gamma_{\text{sp}}} , \quad
f_t = f_\mathrm{eq} , 
\ee
one obtains the result for the GW phase $\Phi(f)$ for the IMRI in a static DM spike [Eq.~\eqref{eq:rho-init}] at Newtonian order.
The Newtonian-order GW phase for the IMRI in a dynamic DM spike was obtained by choosing the parameters to be 
\begin{subequations}
\ba
\eta &=& \frac{5}{8-\gamma_{\text{sp}}}\left(\frac{f_\mathrm{eq}}{f_b}\right)^{(11-2\gamma_{\text{sp}})/3}\,, \\
\lambda &=& \frac{6-2\gamma_{\text{sp}}}{3}\,, \\
\vartheta &=& 1\,,  \\
f_t&=& f_b\,,
\label{Dynamic Phase Parameters}
\ea
\end{subequations}
An ansatz for the break frequency $f_b$ was proposed in~\cite{Coogan:2021uqv} to be 
\be
f_b = \sigma\left(\frac{m_1}{1000M_\odot}\right)^{-\alpha_1}\left(\frac{m_2}{M_\odot}\right)^{\alpha_2}\left(1+ \zeta\log \frac{\gamma_{\text{sp}}}{\gamma_r}\right) ,
\ee
and the five free parameters in this expression were fixed by fitting the GW phase to eighty simulations run by the \textsc{HaloFeedback} code.
The values of these best-fit parameters are $\alpha_1 = 1.4412$, $\alpha_2 = 0.4511$, $\sigma=0.8163$Hz, $\zeta = -0.4971$, and $\gamma_r = 1.4396$. 
For further discussion of the form of this ansatz for the fitting function, see~\cite{Coogan:2021uqv}.

\begin{figure*}[t]
    \centering
    \includegraphics[width=0.49\textwidth]{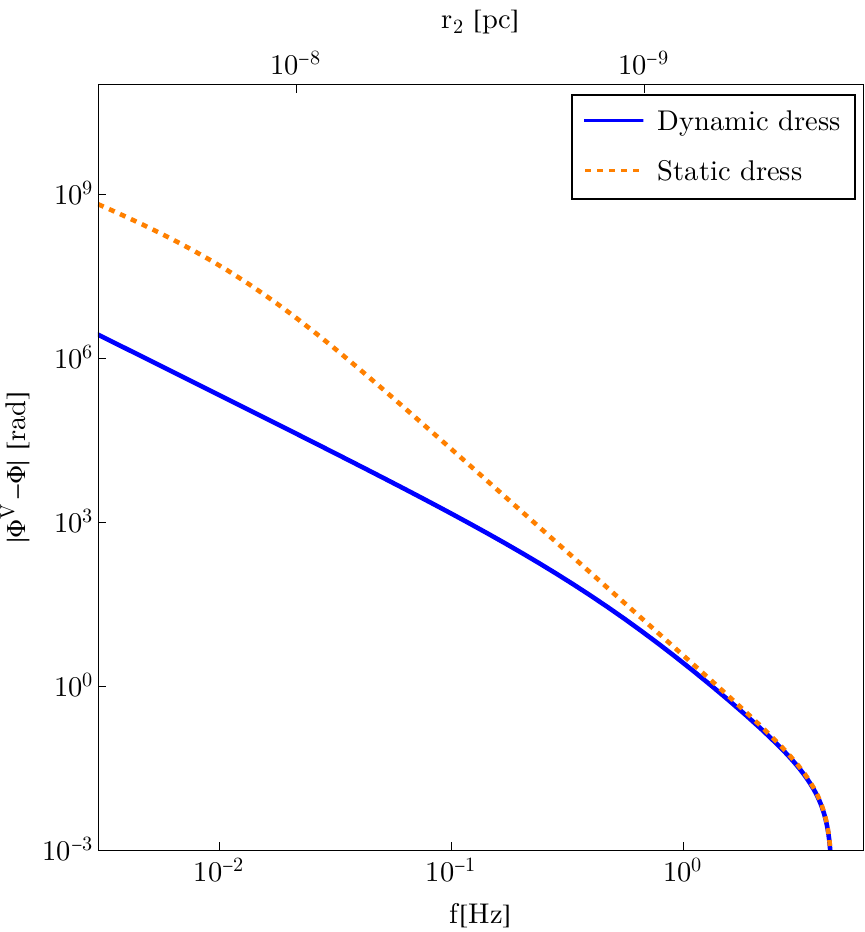}
    \includegraphics[width=0.49\textwidth]{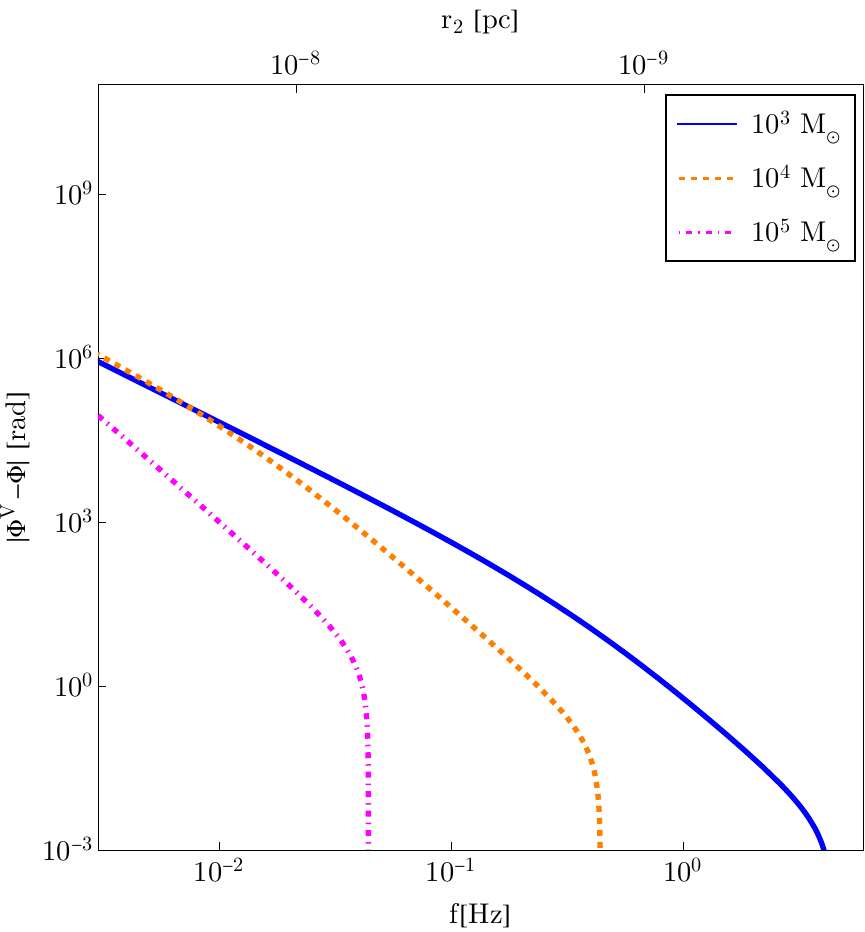}
    \caption{\textbf{Gravitational-wave dephasing for IMRIs with DM from those without}.
    \emph{Left}: A comparison of the GW dephasing for static and dynamic DM spikes versus frequency. 
    The binary masses are given by $(m_1,m_2)=(10^3,1.4)\, \mathrm{M}_\odot$, and the DM parameters are those described in the introduction to Sec.~\ref{Section:GW_SetUp}. 
    The top horizontal axis shows the binary separation $r_2$ which is related to GW frequency $f$ through the Kepler's law $\pi f = \sqrt{(m_1+m_2)/r_2^3}$, for a circular orbit. 
    In both static and dynamic cases, the dephasing resembles a broken power law, with breaks at different frequencies in each case.  
    The inspiral proceeds from left to right and the dephasing goes to zero at the ISCO frequency, because the phases in the vacuum and nonvacuum cases are equated at this value. 
    \emph{Right}: Similar to the left panel but only the dephasing for the dynamic DM case is shown for the three different values of $m_1$ given in the legend.}
    \label{fig:Dephasing} 
\end{figure*}

To illustrate how the dark matter changes the GW phase, we show in Fig.~\ref{fig:Dephasing} the dephasing ($\Phi^V - \Phi$) from a vacuum IMRI with the same parameters of the binary, though without the DM.
The left panel reproduces a similar result in~\cite{Coogan:2021uqv}, and serves as a consistency check.
The right panel shows how the phase differs as a function of different primary masses for a fixed secondary mass.
The dephasing goes to zero at the inner-most stable circular orbit (ISCO) GW frequency, because the vacuum and DM GW phases were chosen to be equal at that value.
Frequencies above that of the ISCOs are not shown, as the phase is valid just for the inspiral (and the plunge and merger would need to be modeled to compute the phase at these higher frequencies).

\subsection{Gravitational-wave phase in the frequency domain} \label{Section:GW_F_Domain}

We next discuss the form of the GW strain in frequency domain.
Because we will be working with the LISA noise curves that are averaged over the sky-position and polarization angles, we will also average the amplitude of the GW strain over these angles as well as the inclination angle and reference phase angle.
This angle averaging removes the dependence of the strain $\tilde h$ on the antenna-response functions $F_+$ and $F_\times$.
When we compute the noise-weighted inner-products of two signals, the factors arising from angle averaging the signals will also cancel with equivalent factors in angle averaging the noise curve $S_n(f)$, as described in more detail in~\cite{Robson:2018ifk}.
Thus, we will drop these factors from the sky-position and polarization average from our waveform and denote this normalized, angle-averaged strain by just $\tilde h(f)$.
It is common then to evaluate the Fourier transform in the stationary-phase approximation (SPA) and to write $\tilde h$ as
\begin{equation}
    \tilde h(f) = A f^{-7/6} e^{i \Psi(f)} .
\end{equation}
Because the phase $\Phi(f)$ was computed using Newtonian physics, we use just the leading, quadrupole PN vacuum term for the amplitude.
The expression for this normalized, angle-averaged amplitude is
\begin{equation}
\label{eq:amp}
    A = \frac{\mathcal{M}^{5/6}}{\sqrt{6} \pi ^{2/3} D_L}\,,
\end{equation}
where $D_L$ is the luminosity distance to the source.\footnote{The numerical factor of $\sqrt{1/6}$ in $A$ comes from the product of two square roots: $\sqrt{5/24}$ and $\sqrt{4/5}$. 
The factor of $\sqrt{5/24}$ is the square root that appears in the amplitude of the plus and cross polarizations of the strain (see, e.g., Eq.~(20) of~\cite{Robson:2018ifk}).
The factor of $\sqrt{4/5}$ comes from the average over the inclination angle and reference phase (see, e.g., Eq.~(16) of~\cite{Robson:2018ifk}). 
We also do not include the DM-induced corrections to the amplitude (so that it has the same form as that in GR), because the amplitude corrections are not as significant as those in the phase (see, e.g.,~\cite{Tahura:2018zuq}).}

Next, we turn to the frequency-domain SPA phase.
The phase $\Phi(f)$ in Eq.~\eqref{GeneralPhaseEq} is the time-domain phase written in terms of frequency---i.e., $\Phi[t(f)]$. 
This differs from the SPA phase $\Psi(f)$ in frequency domain~\cite{Cutler:1994ys}, but the two are related by 
\begin{equation}
\label{eq:phase_freq}
    \Psi(f) = 2 \pi f t(f) - \phi(f) - \frac{\pi}{4} .
\end{equation}
Here we have defined
\begin{equation} \label{eq:phi-of-f}
    \phi(f) = 2 \pi \int^f f \frac{dt}{df} df = - \Phi[t(f)] ,
\end{equation}
which has a relative minus sign from the phase $\Phi(f)$ in Sec.~\ref{Section:GW_T_Domain}.\footnote{The phases $\phi$ and $\Phi$ differ by a minus sign because the latter is defined by~\cite{Kavanagh:2020cfn,Coogan:2021uqv}
\begin{equation}
    \Phi(f) = 2 \pi \int_f f \frac{dt}{df} df . 
\end{equation}
}  
To obtain $\Psi$, we need to compute $t(f)$. 
It can be obtained from differentiating and then integrating Eq.~\eqref{eq:phi-of-f}:
\begin{equation}
      t(f) = \frac{1}{2\pi}\int^f \frac{1}{f} \frac{d\phi}{df}df .
\end{equation}
From the time-domain phase in Eq.~\eqref{GeneralPhaseEq}, we find the frequency-domain phase for the dynamic DM case is given by
\begin{align}
\label{eq:Psi}
    \Psi = {} & \Psi^V(f)- \frac{5\eta}{16(3
   \bar{\lambda }^2+\bar{\lambda }-2)}(\pi \mathcal M f_t)^{-5/3}
y^{-\bar{\lambda }} \nonumber \\
& \times \left[ \,
   _2F_1\left(1,\frac{3\left(\bar{\lambda
   }+1\right)}{5};\frac{3 \bar{\lambda
   }+8}{5};-y^{-5/3}\right) \right. \nonumber \\
&  -\frac 35 \left(\bar{\lambda
   }+1\right)  y^{5/3} \log
   \left(y^{-5/3}+1\right)+ \frac 35
   \bar{\lambda }- \frac 25 \Bigg] .
\end{align}
The vacuum phase, to leading PN order, is given by
\begin{equation}
\label{eq:Psi_V}
  \Psi^V(f) =  2\pi f t_c - \phi_c - \frac{\pi}{4} + \frac{3}{128} (\pi \mathcal{M} f)^{-5/3} , 
\end{equation}
and we have defined
\begin{equation}
    \bar\lambda = \lambda + \frac{5}{3} = \frac{11-2 \gamma_\mathrm{sp}}{3} .
\end{equation}
The parameters $t_c$ and $\phi_c$ are the time and phase of coalescence, respectively.

\subsection{Parameterized post-Einsteinian framework} \label{Section:PPE}

We next review the parameterized post-Einsteinian framework, which was originally developed for testing GR with GW observations with a theory-agnostic approach~\cite{Yunes:2009ke,Yunes:2013dva} (in the sense that the ppE framework was designed to make non-restrictive assumptions about the gravitational-waveform model used to analyze a given GW event, and to permit the GW observation to determine if the ppE hypothesis or the GR hypothesis is favored). 
We will apply this framework in this paper to determine how well the ppE framework can reduce potential biases in parameter estimation and detect (or constrain) the DM effects in GWs from an IMRI immersed in an evolving DM spike.

The ppE template waveform for the $\ell=2$, $m=2$ harmonic that we will use is given by
\begin{equation}
 \tilde h(f) = A f^{-7/6}e^{i(\Psi^{V} + \beta u^b)} , 
\label{Template Waveform}  
\end{equation}
where $u = (\pi \mathcal M f)^{1/3}$, $\beta$ is the ppE parameter that controls the overall magnitude of the DM effects (in the context of this paper), and $b$ is the exponent that is related to the ``effective'' PN order $N$ of the correction term relative to $\Psi^V$ by $b = 2 N-5$.
As we will discuss in more detail below, the phase in Eq.~\eqref{eq:Psi} is not the vacuum GR phase plus a single power law in frequency as in Eq.~\eqref{Template Waveform}, but there is a choice of the parameter $b$ which will minimize errors, which will determine the effective PN order.

\section{Data analysis methods and measurement errors}\label{Section:Fisher}

In this section, we review the data-analysis methods that we use to compute the signal-to-noise ratio (SNR), the systematic errors and the statistical errors.
We use matched filtering and Fisher-matrix methods~\cite{Finn:1992wt,Cutler:1994ys} to compute the SNR and statistical errors, respectively.
In the context of GW data analysis, the inverse Fisher matrix is most naturally interpreted as a Bayesian uncertainty of the posterior probability distribution for a single GW event under several assumptions~\cite{Vallisneri:2007ev}: 
for a high SNR signal with Gaussian noise, when the prior probabilities are constant over the relevant parameter space, the inverse Fisher matrix is the covariance matrix for the true source parameters.

As we show in more detail in Fig.~\ref{fig:SNR}, and describe in Sec.~\ref{subsec:SNR}, the SNR of our signals will approximately 17, which is the SNR threshold for semi-coherent searches derived in~\cite{Chua:2017ujo} (see also~\cite{Gair:2004iv}).
This reasonably high threshold, in addition to the well constrained binary parameters in IMRIs and the reduced parameter set in our angle-averaged waveforms allows the assumptions underlying the Fisher-matrix formalism to be satisfied to a reasonable approximation.
Because we are using a ppE waveform model rather than the exact dynamic DM phase, there will be biases (systematic errors) that arise from using this simpler waveform family.
We estimate these systematic errors using the method described in~\cite{Cutler:2007mi}, which we describe in more detail in Sec.~\ref{subsec:errors}.

\subsection{Signal-to-noise ratio} \label{subsec:SNR}

To compute the SNR, we first define the noise-weighted inner product
\be \label{inner product}
(a|b) = 2\int_{f_\mathrm{low}}^{f_\mathrm{high}}\frac{\tilde a^* \tilde b + \tilde a \tilde b^*}{S_n(f)}df .
\ee
Here $a$ and $b$ represent time series, and $\tilde a$ and $\tilde b$ are their Fourier transforms.
The noise spectral density, $S_n(f)$ will be taken to be the sky and polarization angle-averaged one, as given in~\cite{Robson:2018ifk}.

We next discuss how we determine the lower and upper frequency limits of integration for the noise-weighted inner product in Eq.~\eqref{inner product}.
To do so, we begin by noting that the binaries are at sufficiently large distances that we need to take into account the gravitational redshift $z$ of the frequency between the source frame and the LISA detector frame;
similarly, the source-frame chirp mass $\mathcal M$ must be scaled by $1+z$ to convert to the detector-frame chirp mass $\mathcal M_z$ [i.e., $\mathcal{M}_z = (1+z) \mathcal{M}$].
Because we will specify a luminosity distance $D_L$ to the source, we need to choose a cosmology to obtain the redshift $z$ (specifically, we will use the flat Planck 2015 cosmological parameters~\cite{Planck:2015fie}).

We choose the lower frequency to be the frequency from which the IMRI will merge in a four-year period, assuming it inspirals adiabatically on a sequence of quasicircular orbits;
We use the leading Newtonian-order expression for $df/dt$ in vacuum to compute this value, and we specifically use the convenient expression given in~\cite{Berti:2004bd}, which uses an infinite frequency to define the merger:
\be  \label{low frequency equation}
f_{\text{low}} \approx 0.012  \text{Hz}\left(\frac{\mathcal{M}_z}{50  M_\odot}\right)^{-\frac{5}{8}}\left(\frac{T_{\text{obs}}}{
4\text{yr}}\right)^{-\frac{3}{8}} .
\ee
We choose the upper frequency limit to be
\begin{equation}
    f_\mathrm{high} = \min(f_\mathrm{ISCO},1\,\mathrm{Hz}) ,
\end{equation}
where $f_\mathrm{ISCO}$ is the frequency at innermost stable circular orbit, which is also given in~\cite{Berti:2004bd}: 
\be
f_{\text{ISCO}} = \frac{1}{6^{3/2}\pi M_z} .
\ee
Here $M_z \approx m_{1,z}$ is the redshifted total mass of the binary (which is also approximately the shifted primary mass $m_1$).
The factor of 1\,Hz is our choice for the upper frequency cutoff for LISA.

\begin{figure}[t]
    \centering
    \includegraphics[width=\columnwidth]{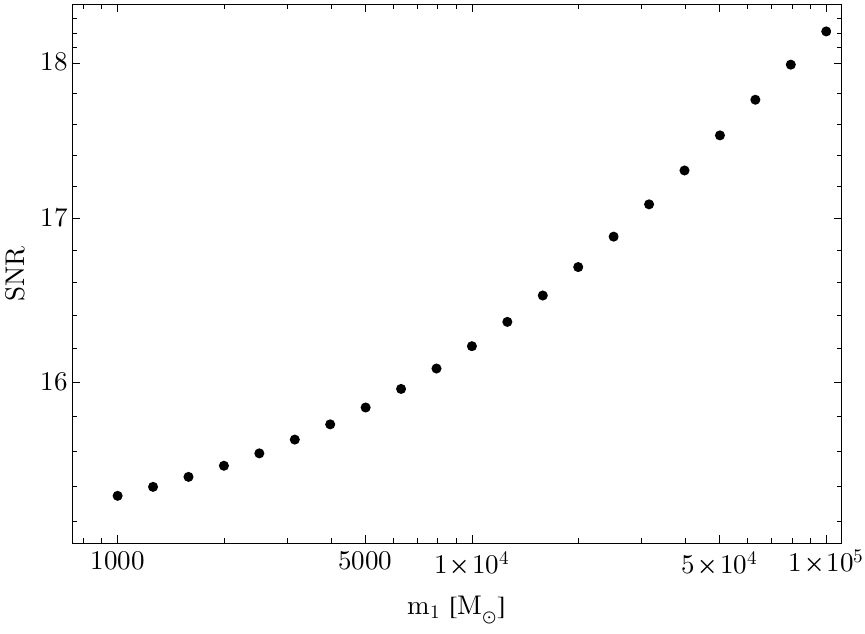}
    \caption{\textbf{SNR as a function of the primary mass}.
    The mass of the secondary has been fixed to $m_2 = 1.4M_\odot$. 
    Our choices of the luminosity distances that produce these SNRs are described in more detail in the text of Sec.~\ref{subsec:SNR}.}
    \label{fig:SNR}
\end{figure}

The SNR is given in terms of the noise-weighted inner product by 
\begin{equation}
    \mathrm{SNR} = \sqrt{(h|h)} .
\end{equation}
The values of the SNR for the IMRIs considered in this paper are shown in Fig.~\ref{fig:SNR}.
In that figure, the primary mass $m_1$ is varied while the secondary mass is fixed to $m_2 = 1.4$\,M$_\odot$. 
To choose the distance, we first computed each SNR with the corresponding source-frame masses and adjusted the luminosity distance to make the SNR equal to 15.
With this luminosity distance, $D_L$ for each binary, we then determined the corresponding redshift via the relationship
\be \label{Redshift Formula}
D_L = \frac{1+z}{H_0}\int_{0}^{z}\frac{dz'}{\sqrt{\Omega_M(1+z')^3 + \Omega_{\Lambda}}} .
\ee
Here $H_0= 67.8$~km/s/Mpc is the local Hubble constant, $\Omega_\Lambda = 0.7$ is the energy density of the cosmological constant (rounded to one significant figure)~\cite{Planck:2015fie}, and $\Omega_M = 1-\Omega_\Lambda$ is the matter density.
We then computed the SNR in Fig.~\ref{fig:SNR} using the redshifted masses.
The distances required to have these SNRs increase monotonically as a function of mass and are approximately 60\,Mpc for the $10^3$\,M$_\odot$ case and 820\,Mpc for the $10^5$\,M$_\odot$ case (rounded to the nearest tens digit).

\subsection{Systematic and statistical errors} \label{subsec:errors}

The systematic error on a given parameter $\theta^j$ is given by~\cite{Cutler:2007mi}
\be \label{SysError_Theta_i}
\Delta_{\text{sys}}\theta^j = \left(\Gamma^{-1}\right)^{jk}\left(i h\Delta\Psi \Big{|}\partial_k h\right), 
\ee
where 
$\Delta \Psi$ is the difference between the phase of the injected (true) and template waveforms.

Computing statistical errors makes use of the Fisher information matrix. 
It is computed from the noise-weighted inner product [Eq.~\eqref{inner product}] of the derivatives of the waveform $h$ with respect to each parameter~\cite{Finn:1992wt,Cutler:1994ys}:
\be \label{BasicFisher}
   \Gamma_{jk} = \left(\pde{ h}{\theta^j}\Big{|}\pde{ h}{ \theta^k}\right) .
\ee
The inverse of the Fisher matrix is the covariance matrix (under the assumption noted above that the priors are constant over relevant parameter range).
However, Gaussian priors centered around the maximum a posteriori values can be incorporated into the formalism.
The standard deviations of these Gaussian priors will be denoted by $\sigma_{\theta_i}^{(0)}$, as in~\cite{Poisson:1995ef,Berti:2004bd,Carson:2020gwh}. 
Assuming that there are no correlations among the prior parameters, one can then define an effective Fisher (inverse covariance matrix $\tilde \Gamma_{ij}$ by 
\be
\tilde{\Gamma}_{ij} = \Gamma_{ij} + \frac{1}{\left(\sigma_{\theta_i}^{(0)}\right)^2}\delta_{ij},
\ee
where $\delta_{ij}$ is the Kronecker delta. 
Thus $(\tilde{\Gamma}^{-1})_{ij}$ is a covariance matrix that takes into account the Gaussian priors.
Our measure of the statistical error for a given parameter $\theta^i$ will be the square root of the diagonal elements of the effective Fisher matrix,
\be
\Delta_{\text{stat}}\theta^i = \sqrt{(\tilde{\Gamma}^{-1})_{ii}} \, .
\ee
The repeated index $i$ is not summed over on the right-hand side.
We only impose a prior on the phase of coalescence $\phi_c$, which is periodic with period $\pi$.
We approximate this in this context by setting $\sigma_{\phi_c}^{(0)} = \pi$. 
The remaining parameters will have no prior imposed, which one can formally describe as setting $\sigma_{\theta_i}^{(0)} = \infty$ for these remaining parameters.

\section{Overview of the ppE Tests} \label{Section:Motivation_Tests}

We give a brief overview of the different studies we perform to assess the performance of the ppE framework for capturing the dephasing effects induced by dark matter and the accuracy with which the vacuum and ppE parameters can be measured.

The first test that we perform examines whether a vacuum merger template is capable of accurately estimating the vacuum parameters for a GW waveform from an IMRI with a dynamic DM spike. 
The work in~\cite{Coogan:2021uqv} examined this question, and determined that the inferred value of the chirp mass could be significantly biased, and the bias increased as the characteristic density of the spike became larger.
The method used in~\cite{Coogan:2021uqv} involved solving an optimization problem that became difficult to solve for large biases.
Thus, it was unable to compute an estimate of the bias for the initial conditions of the DM spike that we treat in this paper.
The method of computing the bias in~\cite{Cutler:2007mi}, however, uses the true parameters of the IMRI waveform (which we assume are known), which allows us to obtain an estimate of the systematic error.
For simplicity, we refer to this test as Question 1 (Q1).

The second test uses the ppE waveform templates under the assumption that the parameter $\beta$ has zero as its true value, but can be nonzero to accommodate waveforms that deviate from vacuum GR ones.
We will compute the statistical and systematic errors for the vacuum waveform parameters and $\beta$.
A nonzero systematic error on $\beta$ that is larger than the statistical error would be an indication that a nonzero ppE phase term would be needed to perform unbiased data analysis.
In this test, we compute the systematic errors for a large number of fixed ppE exponent parameters $b$, and we choose the best one by minimizing the systematic error on the vacuum parameters (particularly, the chirp mass $\mathcal M_z$).
An illustration of this procedure and further details about it are provided in Appendices~\ref{app:analytic} and~\ref{AppendixA}; specifically, we also demonstrate the validity of this procedure by recovering the value of $b$ in the case of a static spike, which is known analytically.
Because we assume $b$ is fixed in each individual computation of the systematic error on the vacuum parameters, we do not include $b$ as a waveform-model parameter which has systematic or statistical uncertainties.
This test will be called Q2.

Our third test involves computing the systematic and statistical errors after shifting the parameters by amounts that are equal to the systematic errors computed in Q2.
We then evaluate the systematic and statistical errors with these updated ``best'' parameter values.
We check to see if the statistical errors are larger or smaller than the systematic errors.
If statistical errors dominate, then the ppE waveform would not have significant bias; if the systematic ones do, then the ppE waveform would require further modification to robustly capture the GW effects of a dynamic DM spike.
As in Q2, we determine $b$ through the procedure described in Appendix~\ref{AppendixA}, and we assume that the optimized $b$ is a fixed parameter with no uncertainties.
This test is Q3.

Our fourth and final test is similar to Q3, but it allows the ppE exponent $b$ to have uncertainty once it is fixed to the value used in Q3.
In this case, we compute the systematic and statistical errors on all parameters in Q3 and also for $b$.
The interpretation of this test, Q4, is the same as in the prior one (Q3). 

For convenience, we summarize the main questions that the four tests are aimed to assess in the list below:
\begin{enumerate}
    \item[Q1.] How large is the bias on the vacuum parameters when using a vacuum template to analyze a GW signal with dynamic DM effects?
    \item[Q2.] Does adding a ppE phase term to the vacuum template reduce bias when the ``fiducial'' value of the ppE phase parameter $\beta$ is assumed to vanish---and if so, by how much?
    \item[Q3.] How accurate is the ppE template when the fiducial parameter values are adjusted to account for the parameter biases, assuming the ppE exponent $b$ has no uncertainty? 
    \item[Q4.] How accurate is the ppE template at capturing DM effects in the gravitational waveform when the parameter $b$ has uncertainty?
\end{enumerate}

\section{Results of the ppE tests}\label{Section:Results}

We now present the results of the four tests that address the questions listed in Sec.~\ref{Section:Motivation_Tests}. 
We assume a four-year observation period with LISA for all the results below.

\subsection{Q1.\ Parameter bias from using vacuum waveforms}\label{test1}

We begin by addressing Q1, which pertains to the amount of bias that arises from using vacuum GW templates to model IMRIs in a dynamical DM spike. 
In this case, the difference in the frequency-domain phase between the simulated signal and template is
\begin{equation}
    \Delta \Psi = \Psi - \Psi^V ,
\end{equation}
where $\Psi$ and $\Psi^V$ are given in Eqs.~\eqref{eq:Psi} and~\eqref{eq:Psi_V}, respectively. 
We choose the vacuum parameters for our study to be
\begin{equation}
    \theta^i = (\ln \mathcal{M}_z, t_c, \phi_c, \ln A)\,,
\end{equation}
with $A$ defined in Eq.~\eqref{eq:amp}.
The true (injected) signal has the time and phase of coalescence set to be $t_c^\mathrm{(fid)} = \phi_c^\mathrm{(fid)} =0$. 
For the chirp mass, we use $m_2^\mathrm{(fid)} = 1.4$\,M$_\odot$ in the source frame, and we vary the value of $m_1$ for the primary black hole. 
The luminosity distances (and thus redshifts) are the same as those used to produce Fig.~\ref{fig:SNR}.

\begin{figure}[t]
    \centering
    \includegraphics[width=\columnwidth]{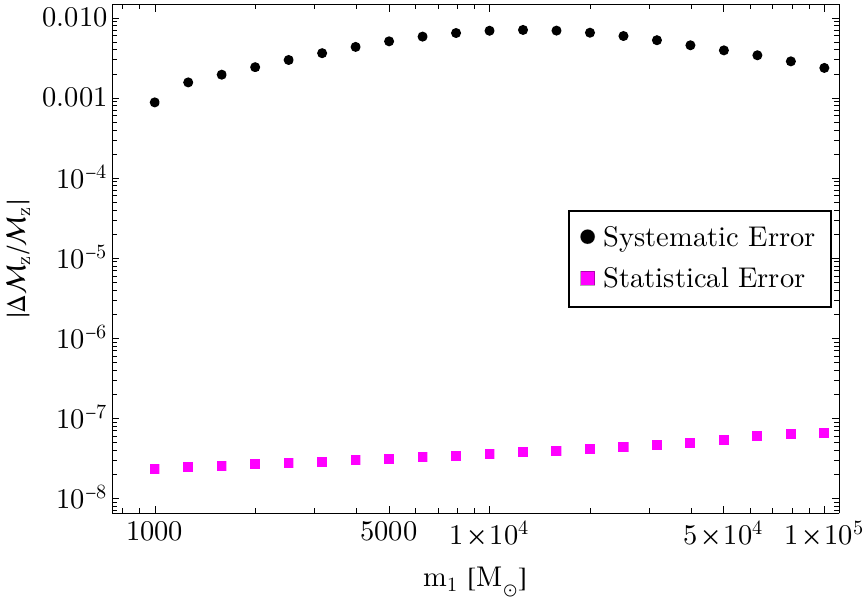}
    \caption{\textbf{(Q1) The relative systematic and statistical errors on the chirp mass against the primary BH mass}.
The errors were computed using a vacuum IMRI template assuming the true signal was an IMRI with a dynamic DM spike. 
The systematic error is significantly greater than the statistical error, which demonstrates the large bias that would arise if one were to attempt to use a vacuum template to analyze GWs from IMRIs in a DM spike.}
    \label{fig:Q1 Results}
\end{figure}

Figure~\ref{fig:Q1 Results} presents the fractional systematic and statistical error on the detector-frame chirp mass as a function of $m_1$. 
Note that the systematic error is consistently higher than the statistical error by at least four orders of magnitude, which is a significant parameter-estimation bias. 
The bias on the detector-frame chirp mass is the most pronounced bias, but the other parameters also have nontrivial biases, which we do not show.

The results above show that if a GW signal from an IMRI within a DM spike is detected through a matched-filtering search against vacuum templates, there will be large biases on the chirp mass. 
Given the results in Fig.~6 of~\cite{Coogan:2021uqv}, however, it will likely be challenging to detect such systems with vacuum templates.
Thus, if a vacuum template is not capable of detecting the signal, then we would assume
that the signal was detected by some other means (such as a coherent excess-power search) and was analyzed with vacuum templates.
%Given these very large biases on the chirp-mass parameter estimation, it is unlikely that a vacuum template would be 
%capable of detecting an IMRI within a DM spike using a matched-filtering search, 
%Thus, such a study implicitly assumes that the signal was detected by some other means (such as a coherent excess-power search) and was analyzed with vacuum templates.

Also note that we do not attempt to determine the source-frame chirp mass, as this would require measuring the luminosity distance (and thus the redshift).
Because the statistical error of such measurements is not nearly as high as that for the chirp mass, propagation of errors would lead to a much larger statistical error on the source-frame chirp mass than that shown in Fig.~\ref{fig:Q1 Results}.

\subsection{Q2.\ Using ppE templates to reduce bias}\label{test2}

Next, we address Q2, which relates to whether analyzing the event with a ppE template leads to less biased results than using vacuum templates (as in Q1). 
The phase difference $\Delta \Psi$ will now include a ppE term,
\begin{equation}
\Delta \Psi = \Psi - \Psi_V - \beta u^b . 
\label{Delta Psi}
\end{equation}
When we compute the Fisher matrix and parameter bias, we now include one additional parameter, the ppE parameter $\beta$:
\begin{equation}
    \theta^i = (\ln \mathcal{M}_z, t_c, \phi_c, \ln A,\beta) .
\end{equation}
The fiducial values for the vacuum parameters are the same as those in Q1 (Sec.~\ref{test1}) and the value of $\beta$ is $\beta^\mathrm{(fid)}=0$. 
For the exponent $b$ [or the effective PN order $N = (b+5)/2$], we determine the best fit value by minimizing the systematic errors on some of the vacuum parameters, as described in Appendix~\ref{AppendixA}.
These minimum-error values of $b$ depend upon the value of the mass $m_1$. 
For example, for $m_1 = 10^3M_\odot$, we find that $N = -0.7121$ gives the minimum systematic error on the chirp mass.

\begin{figure}[t]
    \centering
    \includegraphics[width=\columnwidth]{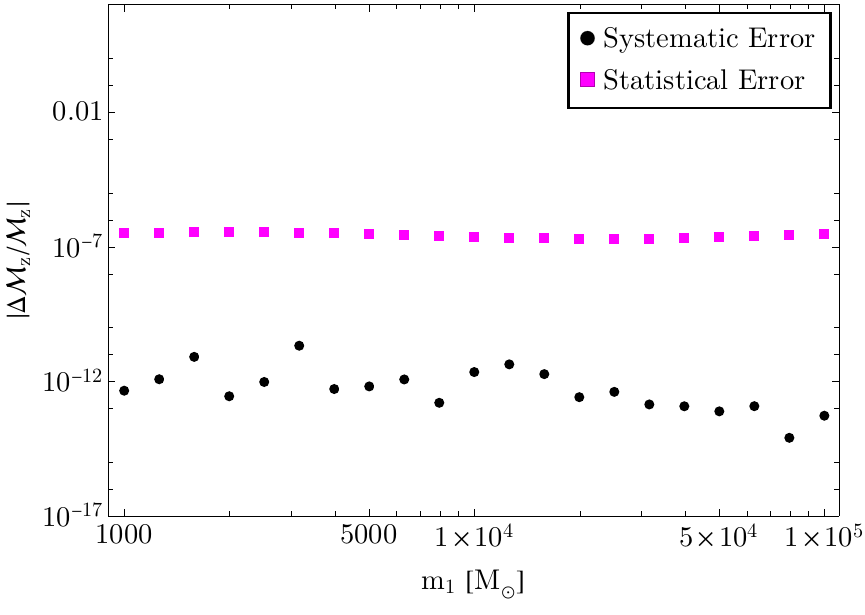}
    \includegraphics[width=\columnwidth]{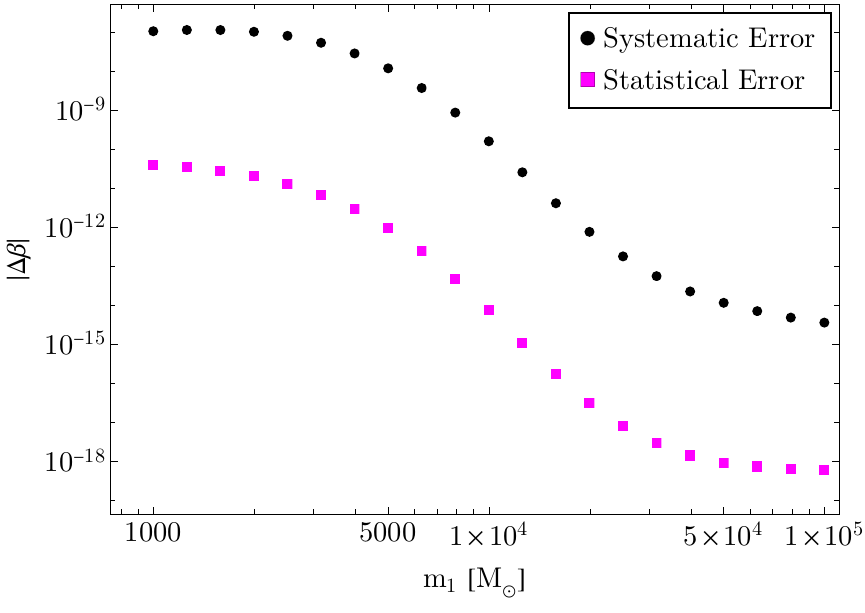}
    \caption{\textbf{(Q2) Errors with a ppE template}.
    \emph{Top}: Errors on the detector-frame chirp mass, similar to those shown in Fig.~\ref{fig:Q1 Results}, but now a ppE template is used and $\beta$ is included in the parameter set. 
    The systematic error has significantly decreased, which indicates that the ppE template can reduce the bias in the parameter estimation from an IMRI with a DM distribution.
    \emph{Bottom}: The absolute systematic and statistical errors for the parameter $\beta$.
    The systematic error is several orders of magnitude larger; thus, the parameter estimation for $\beta$ is significantly biased.}
    \label{fig:Q3 Mchirp Results}
\end{figure}

The top panel of Fig.~\ref{fig:Q3 Mchirp Results} shows the systematic and statistical error on the chirp mass with the ppE phase included in the templates. 
For all tested primary BH masses, the systematic errors are reduced by 8--9 orders of magnitude from the errors for the corresponding cases for Q1 (shown in Fig.~\ref{fig:Q1 Results}). 
The statistical errors, however have increased only by roughly an order of magnitude, which occurs because of the correlation between the chirp mass and ppE parameter $\beta$. 
Overall, the systematic error is consistently smaller than the statistical one by many orders of magnitude, which is an indication that the ppE template could, in principle, remove the bias in the vacuum parameter estimation for GWs from binary black holes in a DM halo.
We only show the result for the chirp mass here, but we have checked that there are also similar decreases in the errors for the other vacuum parameters in the phase.

The bottom panel of Fig. \ref{fig:Q3 Mchirp Results} shows the systematic and statistical error for $\beta$. 
In this case, the systematic error is much larger than the statistical one; this occurs because the effects of DM on the GW phase are captured mainly by the ppE term in the ppE template, thereby producing a large deviation in the estimate for $\beta$ from its fiducial value of zero (see Appendix~\ref{app:analytic} for further details). 
We also note that the statistical error on $\beta$ for $m_1 = 10^3 - 10^4M_\odot$ is roughly consistent with $\Delta \beta \sim 10^{-12}$ for intermediate-mass-ratio inspirals found in~\cite{Chamberlain:2017fjl}.

\subsection{Q3.\ Updating the fiducial values in the ppE templates} \label{test3}

The results in Q2 showed a large systematic error in $\beta$, because we used a fiducial value of $\beta=0$. 
We can rerun our calculations by updating the fiducial values to 
\begin{equation}
    \theta^{i\, \mathrm{(fid, \, new)}} = \theta^{i\, \mathrm{(fid, \, old)}} + \Delta_\mathrm{sys}\theta^i ,
\label{eq:fid_updated}
\end{equation}
where $\theta^{i\, \mathrm{(fid, \, old)}}$ and $\Delta_\mathrm{sys}\theta^i$ are the fiducial values assumed in Q2 and the systematic errors are those determined in Q2. 
Gravitational waveforms with the above updated fiducial values mimic more closely the ``true'' gravitational waveform from an IMRI with a DM halo. 
We keep the value of $b$ to be the same as that used in Q2.

\begin{figure}[t]
    \centering
    \includegraphics[width=\columnwidth]{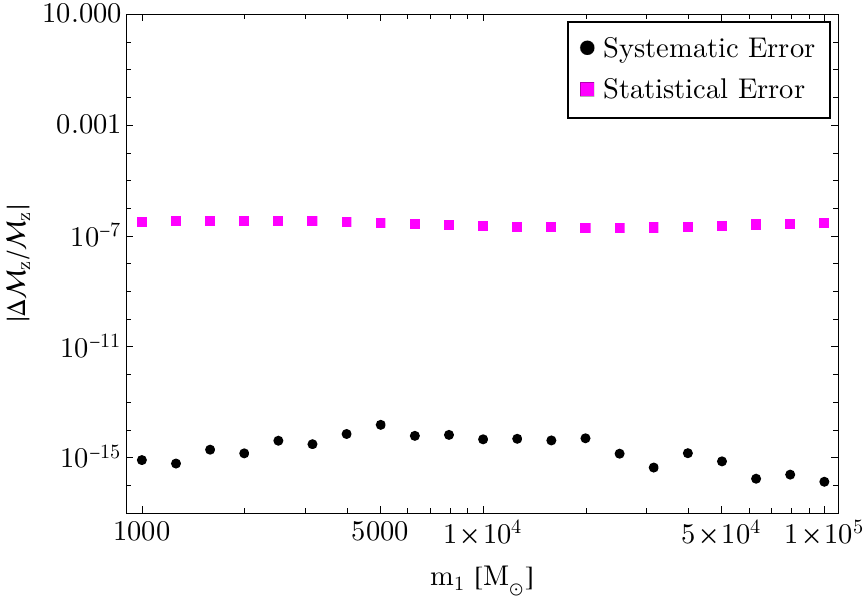}
  \includegraphics[width=\columnwidth]{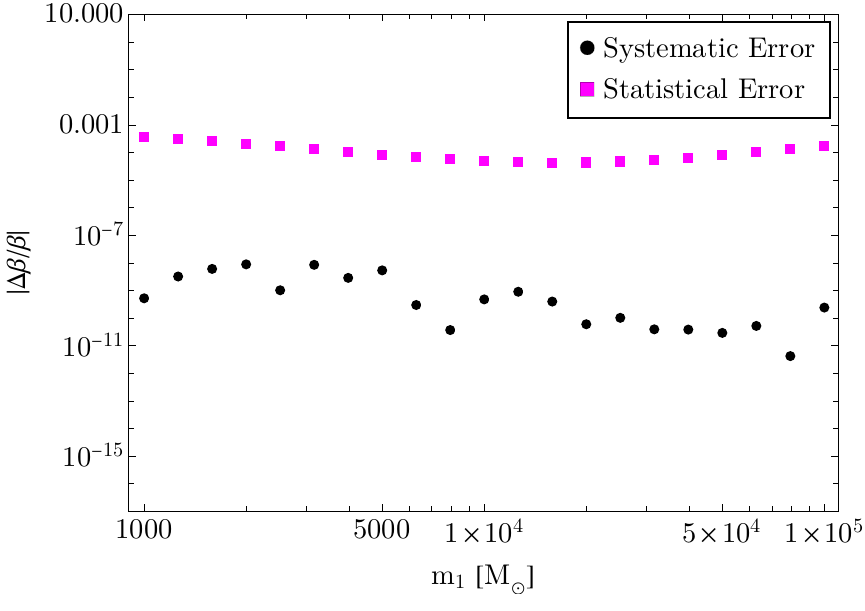}
     \caption{\textbf{(Q3) Errors with updated fiducial parameter values}.
     \emph{Top}: Similar to the top panel in  Fig.~\ref{fig:Q3 Mchirp Results}, but the fiducial values were updated as in Eq.~\eqref{eq:fid_updated}. 
     \emph{Bottom}: Similar to the top panel, but for the ppE parameter $\beta$.
     Note that unlike the bottom panel of Fig.~\ref{fig:Q3 Mchirp Results}, we plot the \emph{fractional} error on $\beta$ instead of the absolute error. In both the top and bottom panels, the systematic errors for both the chirp mass and ppE $\beta$ are much smaller than their respective statistical errors.}
    \label{fig:Q4 Results Mchirp}
\end{figure}

Figure~\ref{fig:Q4 Results Mchirp} shows the errors on the chirp mass and ppE $\beta$ with the updated fiducial values. 
For the chirp mass, the statistical error remains almost unaffected from the results in Q2, while the systematic error has been reduced by a few orders of magnitude. 
For ppE parameter, the fractional systematic error is $\mathcal{O}(10^{-10})$ in most cases; however, the fractional statistical error is $\mathcal{O}(10^{-4})$. 
The fractional statistical error is roughly consistent with the ratio of the absolute statistical and systematic errors in Q2.
It is not precisely the same, however, because the partial derivative $\partial \tilde h/\partial \mathcal M_z$ in the inverse Fisher matrix now contains contributions from terms proportional to $\beta$, which is assumed to be nonzero.
This will introduce additional correlations between $\mathcal{M}_z$ and $\beta$, which would vanish when the fiducial value of $\beta$ is set to be zero, as in Q2. 
Nevertheless, the systematic errors are now much smaller than the statistical errors for both parameters. 
This shows that, in principle, the ppE template is able to reduce the bias in vacuum parameters and to capture the effects of a dynamical DM distribution with a nonzero fiducial value of $\beta$.

\subsection{Q4.\ Uncertainties in the ppE exponent $b$}\label{test4}

In the previous two tests, we determined the optimal value of ppE exponent $b$ that minimized the systematic error on the chirp mass, and we assumed that it was known without any uncertainty. 
This followed one viewpoint on the ppE framework, in which $b$ is not considered a free-parameter of the framework; rather the ppE tests are run with fixed $b$ for the values appropriate for a few common, discrete post-Newtonian orders at which different deviations from GR could conceivably occur (based on specific examples of deviations that arise in theories of gravity that contain additional interactions not present in GR).
However, the dark-matter effects can produce a dephasing that (at least locally in the frequency domain) would be best fit by a $b$ parameter that is not constrained to be any particular PN order.
The power law $\gamma_\mathrm{sp}$ in the initial DM density in Eq.~\eqref{eq:rho-init} can take on a continuum of values in an interval on the real axis, and the best-fit $b$ parameter in these cases would also be a continuum on another real interval, rather than at a few discrete values at a few distinct PN orders.

Thus, in the context of capturing dynamical DM effects, it is more natural to consider $b$ to be a free parameter in the ppE framework, which would be determined simultaneously with the other parameters and would have systematic and statistical uncertainties.
This then leads to the final test, Q4, which addresses how the results in Q3 change if $b$ is included in the parameter set:
\begin{equation}
    \theta^i = (\ln \mathcal{M}_z, t_c, \phi_c, \ln A,\beta,b) .
\end{equation}
We thus will fix the fiducial values for the vacuum parameters and the ppE parameter $\beta$ to be the same as those in Q3.
The fiducial values for $b$ for different $m_1$ are the same as those used in Q2 and Q3.

\begin{figure*}[t]
    \centering
    \includegraphics[width=\linewidth]{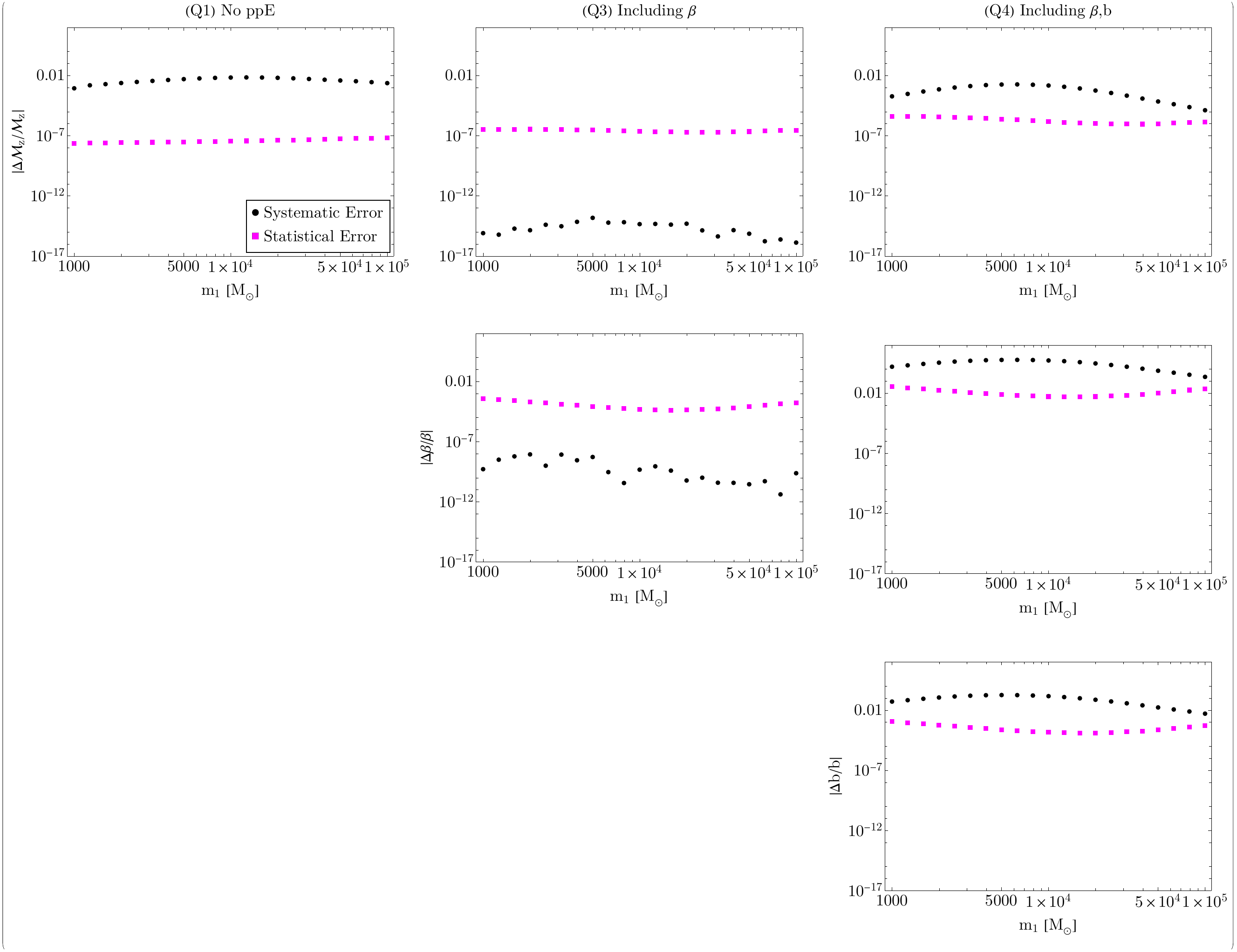}
     \caption{ \textbf{Comparison of the statistical and systematic errors in the different tests}.
     The left column (Q1) and central column (Q3) reproduce the results in Figs.~\ref{fig:Q1 Results} and~\ref{fig:Q4 Results Mchirp}, respectively.
     They are given to make the comparison with the new results for Q4 in the right column easier.
     The first row allows for a comparison of the statistical and systematic errors on the chirp mass in tests Q1, Q3 and Q4.
     The second row gives a similar comparison for the ppE parameter $\beta$ between Q3 and Q4.
     The third row shows the errors on the parameter $b$, which is computed only in Q4.
     Adding $b$ as a free parameter with errors significantly increases the systematic error compared with other tests, and increases the statistical error by around an order of magnitude.}
    \label{fig:Q1-5 Results}
\end{figure*}

Figure~\ref{fig:Q1-5 Results} summarizes the fractional statistical and systematic errors on the chirp mass, $\beta$, and $b$ in Q4, in the right column.
The results of Q1 in Fig.~\ref{fig:Q1 Results} are included in the left column and those of Q3 in Fig.~\ref{fig:Q4 Results Mchirp} are shown in the middle column (so as to make a comparison of the three tests easier).
By considering $b$ as a waveform-model parameter, the statistical errors on the chirp mass and $\beta$ are about an order of magnitude higher than those in Q3, presumably because of increased correlations with the chirp mass and ppE parameter $\beta$.
The systematic errors are significantly larger than the equivalent ones in Q3; thus $\Delta_\mathrm{sys} \mathcal{M}_z/\mathcal{M}_z$ and $\Delta_\mathrm{sys}\beta/\beta$ now exceed the statistical errors in Q4. 

The reason for the increased systematic error can be understood from considering the linear equations in Eq.~\eqref{SysError_Theta_i}.
For the five-parameter set $\theta^i$ in Q3, the value of $b$ was optimized to make the systematic error (the product of the inverse Fisher matrix with the five-component vector of noise-weighted inner products) small.
In Q4, however, the parameter set contains a sixth component, so the inverse Fisher matrix becomes a six-by-six matrix. 
It multiplies a vector with the same first five components as in Q3, but with a sixth component $(i h\Delta \Psi| \partial h/\partial b)$ added.
There is not a simple relationship between the five-by-five block of the inverse Fisher matrix from Q4 (with the same parameters as those from Q3) and the inverse Fisher matrix in Q3.
Thus, using the optimal values from Q3 in Q4 will not necessarily preserve the small systematic errors (and the results in Fig.~\ref{fig:Q1-5 Results} demonstrate that it does not).
In principle, one should perform a new optimization (potentially over multiple parameters) to see if there exist other combinations of parameters that can reduce the systematic errors in Q4.
However, we were unable to determine such parameters.
Thus, we were not able to use this simplest ppE model to capture DM effects and obtain an unbiased estimate on the chirp mass and $\beta$ parameter.

\section{Conclusions}\label{Section:Conclusions}

In this paper, we investigated IMRI systems within an evolving DM spike, and we examined how well the ppE framework could model the GW effects of the DM in these systems.
We first showed that using a vacuum template to model a merger within a dynamic DM spike causes the parameter estimation to be biased. 
Using vacuum templates to analyze systems with DM is most reasonable if the system is detectable with vacuum templates; however, Ref.~\cite{Coogan:2021uqv} showed that the Bayes factor for the DM versus vacuum hypothesis strongly favored DM systems, thus indicating that vacuum templates would likely not be able to detect these systems for the systems that we considered.
This study then supposed that the GW signal with DM could be detected by some other means.

We next performed additional tests within the context of the ppE framework.
The ppE approach often makes the assumption that the true signal is that of vacuum GR, and it places constraints on a parameter $\beta$ in the GW phase that permits a deviation from GR by computing its statistical error. 
Dark-matter effects in the GW phase, however, do not have a zero coefficient, so this approach would only be unbiased if the statistical error when using the ppE framework exceeded the systematic error from assuming vacuum GR is the true signal.
After optimizing the exponent in the ppE phase, the statistical errors were several orders of magnitude smaller than the systematic ones; thus, a nonzero value of $\beta$ is necessary to have unbiased results.

We then performed a similar study in which the fiducial value of the $\beta$ ppE parameter was assumed to be nonzero.
With an optimized ppE exponent $b$, the statistical errors were no larger than the systematic ones.
This suggested that if the exponent was considered to be fixed (and not a parameter with its own errors) then the ppE framework could be used for parameter estimation without introducing bias.

The systematic errors on the vacuum parameters depend sensitively on the choice of the effective PN order; specifically, if the value is slightly off from the optimized one, the error goes up by several orders of magnitude. 
To run the ppE framework with a fixed exponent $b$, it would be necessary to perform the analysis for a large number of finely sampled $b$ values.
This may not be feasible, so it is more natural to treat the ppE parameter $b$ as an undetermined model parameter that has its own systematic and statistical errors.
Using the same optimized value from the previous test caused the systematic errors to exceed the statistical errors, and attempts at further optimization of the fiducial parameters did not significantly improve the results.
We conclude that although the ppE template can, in principle, remove the bias in the parameter estimation for GWs from binary black holes with a DM halo, it may be difficult in practice to do so using the simplest ppE framework.
%\corr{Our results also suggest that an environmental effect, such as the DM spike around an IMRI studied here, can give rise to a nonvanishing measurement of $\beta$ and contaminate model-independent tests of gravity with GWs, which was the original goal of the ppE framework.}

There are several different avenues for future work. 
One reason that the single-term ppE framework may not work well is that the GWs from IMRIs with evolving DM have deviations from vacuum waveforms that are not a single power law. 
Adding multiple ppE terms could produce a better fit to the phase, if they could be optimized to best match the phase from the IMRIs with DM. 
The case that we studied in this paper used just the leading-order Newtonain, quadurpolar waveform for an IMRI on a circular orbit, with a nonspinning primary BH surrounded by a DM distribution that evolves from the injection of energy from dynamical friction.
Many of assumptions could be generalized.
For example, if the secondary is a BH, then accretion onto the secondary introduces additional dephasing from systems that treat only dynamical friction.
The GW effects enter at a different effective PN order from the dynamical-friction ones~\cite{Nichols:2023ufs}, and it would be interesting to investigate how well the ppE framework could account for both accretion and dynamical-friction feedback.
IMRIs are strongly relativistic systems, so it would be useful to include higher PN terms and additional spherical-harmonic modes of the waveforms.
Understanding how well the ppE framework would perform in this context would also be useful to investigate.

Finally, the intent of the original ppE framework is to test for theory-agnostic deviations from GR that arise at fixed PN orders, most frequently at half-integer values. 
Our results have shown that a nonzero ppE parameter $\beta$ significantly reduces the systematic error in our Fisher-based parameter-estimation studies, which suggests that waveforms with nonzero $\beta$ should be used to analyze the GW signals from such systems.
In the context of the ppE framework, a nonzero $\beta$ is associated with a deviation from GR, which might suggest that an environmental effect from a dynamic DM spike instead could be interpreted as a deviation from GR.
However, the systematic error depended sensitively on the ppE exponent parameter $b$, which took on a continuum of values rather than those associated with a particular half-integer PN order.
The fact the best value of $b$ is not one of these discrete values is a likely indication that the origin of the nonzero $\beta$ is an environmental effect rather than a beyond-GR effect.
However, it may be the case that in practical ppE studies, the tests will be performed only a handful of half-integer PN orders.
As a future study, it would also be interesting to investigate how well parameter estimation with ppE waveforms with these discrete ppE exponents $b$ perform when the waveform of an IMRI with a dynamic DM spike is used as the injected signal.
Such studies would determine whether or not the environmental effects arising from a dynamic DM spike could mimic those effects more commonly associated with gravitational theories beyond GR.

\acknowledgments

D.A.N.\ acknowledges support from the NSF Grant PHY-2309021.
K.Y.\ acknowledges support from NSF Grant PHY-2339969 and the Owens Family Foundation.

\appendix

\section{Analytic calculation of the systematic error}
\label{app:analytic}

When the phase difference from vacuum, $\Delta \Psi$, can be written in a ppE form, $\Delta \Psi = \alpha u^a$ (for some constants $\alpha$ and $a$), then the systematic error computed using the ppE templates has a simple form.
Specifically, if we use the ppE template $h$ in Eq.~\eqref{Template Waveform} to recover the true waveform, then the systematic error on each parameter $\theta^i$ becomes
\begin{equation} \label{testingtheta1}
\Delta_\mathrm{sys}\theta^j = \alpha\Big{(}\Gamma^{-1}\Big{)}^{jk}\Big{(}i u^a h\Big{|}\partial_k h\Big{)} .
\end{equation}
Next, we will choose the ppE exponent in the template to be $b = a$. 
This implies that $\partial_\beta\tilde{h}= i u^a h$, so the above systematic error becomes
\begin{equation}
    \Delta\theta^i =  \alpha\Big{(}\Gamma^{-1}\Big{)}^{ij}\Big{(}\partial_\beta h\Big{|}\partial_j h\Big{)} .
\end{equation}
Using the fact that $(\partial_\beta h | \partial_j h) = \Gamma_{\beta j}$ are the $\beta$ components of the Fisher matrix, we then have

\begin{equation}
\Delta\theta^i = \alpha \left(\Gamma^{-1}\right)^{ij} \Gamma_{\beta j} = \alpha \mathbf{I}^{i\beta} = \begin{cases}
    \alpha & (i=\beta)\\
    0 & (i \neq \beta)
\end{cases} ,
\end{equation}
where $\mathbf{I}$ is the identity matrix. 

The above analytic estimate shows that if the PN order (or the exponent) in the ppE term is chosen appropriately, then under the assumption made on $\Delta \Psi$, the systematic error appears in that of the ppE parameter $\beta$ without affecting estimation of other parameters. 
This, in turn, can be used to determine the effective PN order of the DM effect. 
Namely, one can look at the systematic errors in the other parameters (e.g., $\Delta_\mathrm{sys} \ln \mathcal{M}$) and identify the PN order that minimizes this systematic error.
We illustrate this approach in the next appendix.

\section{Optimal PN orders for dark-matter-induced dephasing} \label{AppendixA}

\begin{figure*}[t]
    \centering
    \includegraphics[width=0.97\columnwidth]{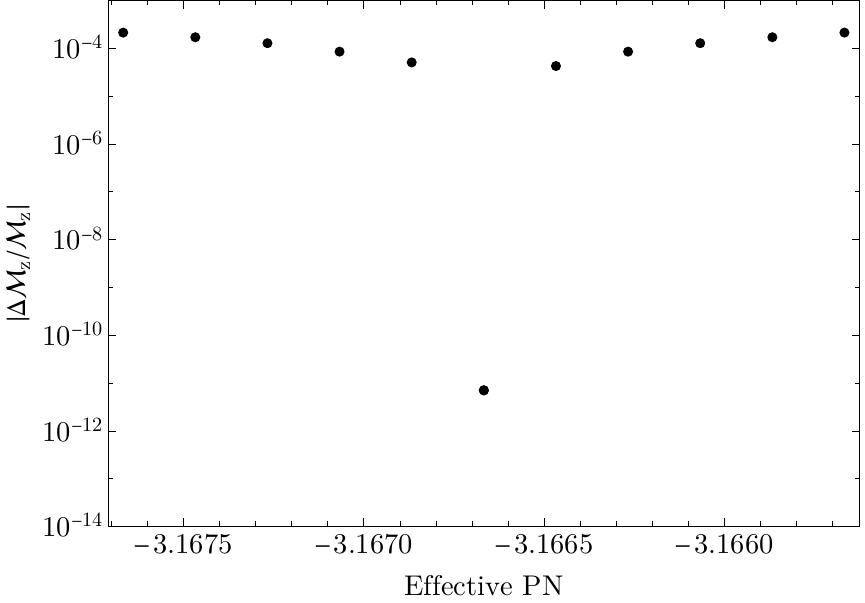}
    \includegraphics[width=\columnwidth]{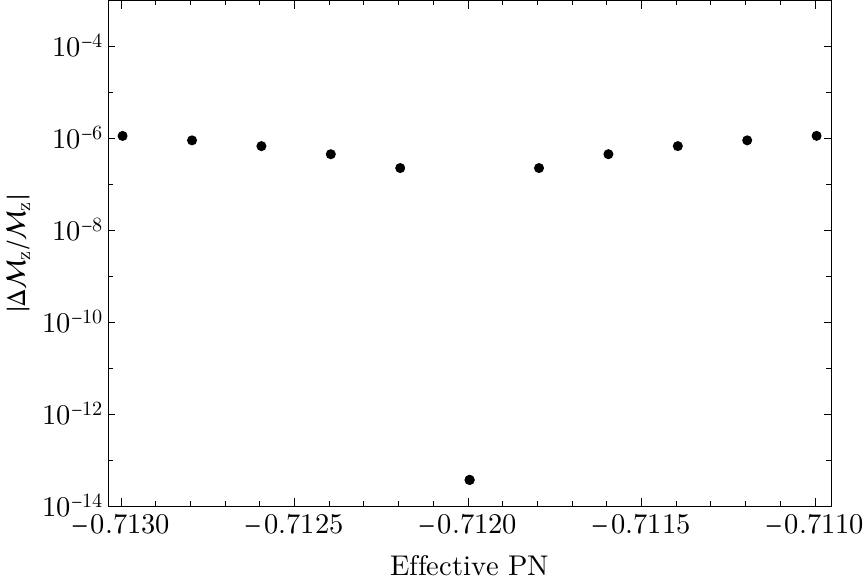}
    \caption{\textbf{Optimizing over different PN orders}.
    \emph{Left}: The fractional systematic error on the chirp mass against the ``effective'' PN order $N = (b+5)/2$ in the ppE term for a static DM distribution. 
    Note the very sharp drop near the PN order of $N = -19/6 = -3.16667$ that can be computed analytically in this case when the DM effects are treated as a perturbation to those of gravitational radiation reaction.
    \emph{Right}: A similar plot as that shown in the left panel, but the waveform phase now comes from the waveform for an IMRI in an evolving DM spike.}
    \label{fig:Finding PN}
\end{figure*}

Here we illustrate how we use the results of Appendix~\ref{app:analytic} to identify the optimal PN-order $N$ for the GW dephasing induced by DM, which minimizes the systematic error on the chirp mass. 
The order $N$ is related to the ppE exponent by $b = 2N-5$.
We illustrate the procedure for both static and dynamic DM spikes.

\subsection{Static DM case}
\label{AppendixB}

We first illustrate the method described in Appendix~\ref{app:analytic} for a static DM distribution. 
For $f \gg f_\mathrm{eq}$, with $f_\mathrm{eq}$ given in Eq.~\eqref{eq:feq}, the hypergeometric function in Eq.~\eqref{GeneralPhaseEq} can be expanded for small $1/y$, and the phase difference $\Delta \Psi$ has the following scaling with frequency:
\begin{equation}
    \Delta \Psi \propto f^{-2(8-\gamma_\mathrm{sp})/3} .
\end{equation} 
For example, when $\gamma_\mathrm{sp} = 7/3$, the above dephasing scales with $f$ as $\Delta \Psi \propto f^{-34/9}$. Comparing this with the ppE term in the phase, we find that the ppE exponent $b$ corresponds to $b = -34/3$, which means that the effective PN order should be $N = -19/6$. 

The left panel of Fig.~\ref{fig:Finding PN} shows the fractional error on the chirp mass as a function of the effective PN order $N$ in the ppE term. 
The template used is the ppE waveform with the fiducial parameter of $\beta =0$, and the masses are chosen to be $(m_1,m_2) = (10^3,1.4)$\,M$_\odot$. 
Notice that the systematic error drops significantly around the PN order of $N = -19/6$ as estimated analytically above, which agrees with our analytic calculation in Appendix~\ref{app:analytic}.
Notice also that the error is very sensitive to the choice of the PN order, in that it rises quickly if the PN order is slightly off from the optimal value. 
We found that we need a precision of at least 5 decimal places in $N$ to find an accurate optimal value.

\subsection{Dynamic DM case}

We next examine at the dynamic DM case which has a more complicated frequency dependence in the phase than the static DM case. 
The right panel of Fig.~\ref{fig:Finding PN} shows the fractional systematic error on the chirp mass for the dynamic DM case. 
Notice that the effective PN order has changed significantly from the static DM case and is now around $N = -0.7121$. 
Once again, the systematic error depends very sharply on the choice of the optimal PN order.

\bibliography{memory}

\end{document}